\documentclass[journal,twocolumn]{IEEEtran}

\usepackage{amsfonts}
\usepackage{times}
\usepackage{graphicx}
\usepackage{latexsym}
\usepackage{dsfont}
\usepackage{amssymb}
\usepackage{amsmath}
\usepackage{cite}
\usepackage{verbatim}
\usepackage{subfigure}

\newcommand{\figref}[1]{{Fig.}~\ref{#1}}

\def\bb0{{\mathbb{0}}}

\def\bb{{\mathbf{b}}}

\def\bff{{\mathbf{f}}}
\def\bg{{\mathbf{g}}}

\def\b0{{\mathbf{0}}}

\def\bX{{\mathbf{X}}}

\def\bbE{{\mathbb{E}}}

\def\sf0{{\mathsf{0}}}

\DeclareMathOperator*{\argmax}{argmax}

\usepackage{epstopdf}
\usepackage{enumerate}
\usepackage{algorithmicx}
\usepackage{algorithm}
\usepackage{amsmath}
\usepackage[noend]{algpseudocode}
\usepackage{float}
\usepackage{hyperref}
\usepackage{color}
\usepackage{makeidx}
\usepackage{bbm}
\usepackage{graphicx}
\usepackage{lipsum}
\usepackage{subfigure}
\usepackage{tablefootnote}
\usepackage{multirow}
\usepackage{multicol}
\usepackage{balance}
\usepackage{booktabs}
\usepackage{soul}
\usepackage[normalem]{ulem}
\usepackage{xcolor}

\newcommand{\sref}[1]{{Section}~\ref{#1}}

\newcommand{\comm}[1]{}

\begin{document}

\title{Vision-Position  Multi-Modal Beam Prediction Using Real Millimeter Wave Datasets \thanks{This work was supported in part by the Air Force Research Laboratory Visiting Faculty Research Program SA2020051003-V0224. Approved for public release AFRL-2021-2366.}}
\author{Gouranga Charan$^{1}$, Tawfik Osman$^{1}$, Andrew Hredzak$^{1}$, Ngwe Thawdar$^{2}$, and Ahmed Alkhateeb$^{1}$\\ $^{1}$Arizona State University - Emails: \{gcharan, tmosman, ahredzak, alkhateeb\}@asu.edu \\ $^{2}$Air Force Research Lab - Email: ngwe.thawdar@us.af.mil}

\maketitle

\begin{abstract}
Enabling highly-mobile millimeter wave (mmWave) and terahertz (THz) wireless communication applications requires overcoming the critical challenges associated with the large antenna arrays deployed at these systems. In particular, adjusting the narrow beams of these antenna arrays typically incurs high beam training overhead that scales with the number of antennas. To address these challenges, this paper proposes a multi-modal machine learning based approach that leverages positional and visual (camera) data collected from the wireless communication environment for fast beam prediction. The developed framework has been tested on a real-world vehicular dataset comprising practical GPS, camera, and mmWave beam training data. The results show the proposed approach achieves more than $\approx$ 75\% top-1 beam prediction accuracy and close to 100\% top-3 beam prediction accuracy in realistic communication scenarios.

\end{abstract}

\section{Introduction} \label{sec:Intro}
Communication over millimeter wave (mmWave) and terahertz (THz) bands is the key enabler for the high data rate requirements in 5G, 6G, and beyond \cite{Rappaport2019}. These systems, however, need to deploy large antenna arrays and use narrow beams at the transmitters and receivers to guarantee sufficient receive signal power. Selecting the optimal beams for these large antenna arrays is typically associated with large training overhead, which makes it difficult for these mmWave/THz communication systems to support highly-mobile wireless applications such as virtual/augmented reality and connected vehicles. This motivates looking for new approaches to overcome this beam training overhead and enable highly-mobile mmWave/THz communication systems. 

Developing solutions for the mmWave beam training and channel estimation overhead has attracted considerable interest over the last decade \cite{Zhang2021a,Alkhateeb2014,Jayaprakasam2017,HeathJr2016,Alkhateeb2018,Khan2020a,Alrabeiah_camera}. These solutions have generally focused on constructing adaptive beam codebooks \cite{Zhang2021a,Zhang2021b}, designing beam tracking techniques \cite{Jayaprakasam2017}, and leveraging the channel sparsity and efficient compressive sensing tools \cite{Alkhateeb2014,HeathJr2016}. These classical approaches, however, can typically save only one order of magnitude in the training overhead, which is not sufficient for large antenna array systems and highly-mobile applications. This motivated developing machine learning approaches that leverage prior observation and side information such as receive signal signature \cite{Alkhateeb2018}, user position \cite{Khan2020a}, and visual/camera images \cite{Alrabeiah_camera} for fast mmWave/THz beam prediction. Each of these sensory data, however, has its own limitations. For example, practical positioning sensors do not normally provide accurate enough positions for narrow beam alignment, and visual/camera data may be sensitive to lighting/weather conditions. 

In this paper, we develop a novel framework for mmWave/THz beam prediction using multi-modal data collected from position and vision (camera) sensors. Thanks to leveraging different data modalities, the proposed approach is more robust to the inaccuracies in practical position sensors and the possible impact of various lighting conditions on the visual information. The proposed approach has been evaluated using a real-world dataset comprising position, vision, and mmWave beam training data to show its potential in reality.

\begin{figure}[!t]
	\centering
	\includegraphics[width=0.85\linewidth]{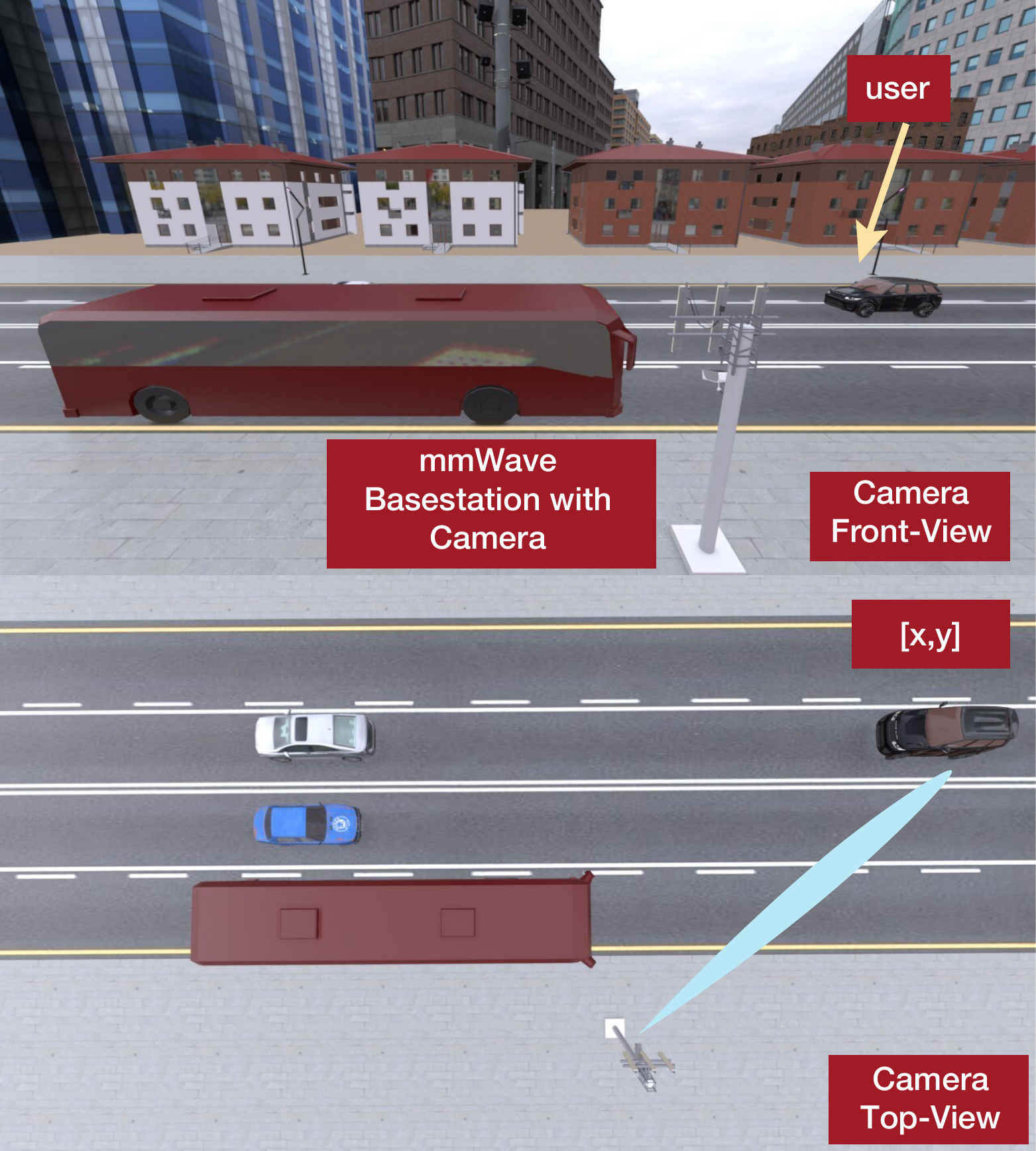}
	\caption{This figure illustrates the considered system and highlights the value of leveraging the positional and visual/camera side information for efficient mmWave/Thz beam prediction.}
	\label{main_fig}
	\vspace{-4mm}
\end{figure}

\section{Millimeter Wave Beam Prediction:\\ System Model and Problem Formulation}
Finding the optimal transmitter/receiver beams is a challenging problem at mmWave and THz communication systems.  This is primarily because of the large control overhead associated with conventional beam training techniques for systems with large antenna arrays \cite{HeathJr2016}. 
In this work, we leverage machine learning and side information collected from visual and position sensors to efficiently predict the beamforming vectors and reduce the beam training overhead. In particular, this work poses the beam prediction task as a multi-modal classification problem: Based on the visual and position data available from the wireless environment, the goal is to predict the best beamforming vector from a pre-defined beam codebook. In the following subsections, we describe the adopted system model and problem formulation before presenting the proposed beam prediction approach in \sref{sec:prop_sol}.

\subsection{System Model} \label{sec:sys_ch_mod}

The considered communication system model consists of a mmWave base station equipped with an $N$-element antenna array and an RGB camera. This base station is serving a  mobile user that is equipped with, for simplicity, a single antenna. The communication system adopts a pre-defined beamforming codebook $\boldsymbol{\mathcal F}=\{\mathbf f_m\}_{m=1}^{M}$, where $\mathbf{f}_m \in \mathbb C^{N\times 1}$ and $M$ is the total number of beamforming vectors in the codebook. For any mmWave user $u$ in the wireless environment with a channel $\mathbf h_{u} \in \mathbb C^{N \times 1}$, if the base station used the beamforming vector $\bff_m$ to serve it, $\bff_m \in \boldsymbol{\mathcal F}$, then the downlink received signal is given by
\begin{equation}
    y_{u} = \mathbf h_{u}^T \bff_m x + n,
\end{equation}
where $x \in \mathbb C$ is the transmitted complex symbol, $\bbE{\left|x\right|^2}=P_T$, with $P_t$ denoting the average transmit power. The variable $n$ represents the receive noise with $n \sim \mathcal N_\mathbb C(0,\sigma^2)$.

\subsection{Problem Formulation} \label{sec:prob_form}
The primary task of this paper is the multi-modal beam prediction for the single user scenario utilizing visual and position information. The task of beam prediction can be defined as determining the optimal beamforming vector $\boldsymbol{\mathbf f^\star}$  out of the candidate beams in the codebook $\boldsymbol{\mathcal F}$, such that the received signal power is maximized.  Mathematically, we can express the beam selection problem as 
\begin{equation}
    \mathbf{f}^{\star} = \argmax_{\mathbf{f} \in \mathcal{F}} \parallel \mathbf h_{u}^T\mathbf{f} \parallel_{2}^{2}.
\end{equation}

In this work, instead of relying on the explicit channel knowledge, i.e., the knowledge of $\mathbf h_{u}$, which is typically hard to acquire in mmWave/THz systems, we target predicting the optimal beam based only on the availability of (i) the RGB images of the camera installed at the base station and (ii) the real-time position information of the user. For that, we define $\bg_u \in \mathbb R^2$  as the two-dimensional position vector (carrying latitude and longitude information) of user $u$ at a certain moment. Further, we define $\bX_u \in \mathbb{R}^{W \times H \times C}$ as the corresponding RGB image of the user $u$, captured by a camera installed at the base station, where $W$, $H$, and $C$ are the width, height, and the number of color channels for the image. 

The objective is then to utilize the input image $\bX_u$ and the location information $\bg_u$ of user $u$ to  predict the optimal beam to serve it, $\mathbf{f}^{\star}$, with high success probability. In this paper, we target developing a  machine learning model to achieve this objective. The model is developed to learn a prediction function $f_{\Theta}(\mathcal S)$, where $\mathcal{S} = (\bX_u, \bg_{u})$ that takes in the observed image-location pair and produces a probability distribution $\mathcal P \in \{p_1, \ldots, p_{M} \}$ over the beams of the codebook $\boldsymbol{\mathcal F}$. The index of the element with maximum probability determines the predicted beam vector. Formally,  
\begin{equation}
\hat{\mathbf f} = \mathbf{f}_{\widehat{m}}, \ \ \ \widehat{m} = \argmax_{m \in \{1, \ldots, M\}} p_m.
\end{equation}
 
This prediction function, $f_{\Theta}(\mathcal S)$, is parameterized by a set $\Theta$ representing the model parameters and learned from a dataset $\mathcal D$ of labeled samples, $\mathcal D = (\mathcal{S}, \mathbf f^{\star})$, consisting of the input image-location pair and its ground truth optimal beamforming vector. Following the formulation convention in machine learning, the objective of the  machine learning model is then to maximize the overall success probability in predicting the optimal beam. Mathematically, we can write 
\begin{equation}\label{joint_prob}
    f^\star_{\Theta^\star}= \underset{f_{\Theta}(\mathcal S)}{\text{argmax}} \  \mathbb{P}(\hat{\mathbf{f}} = \mathbf{f}^{\star}|\mathcal{S})
\end{equation}

Next, we present our proposed multi-modal machine learning model for mmWave/THz beam prediction. 

\begin{figure*}[t]
	\centering
	\includegraphics[width=0.9\linewidth]{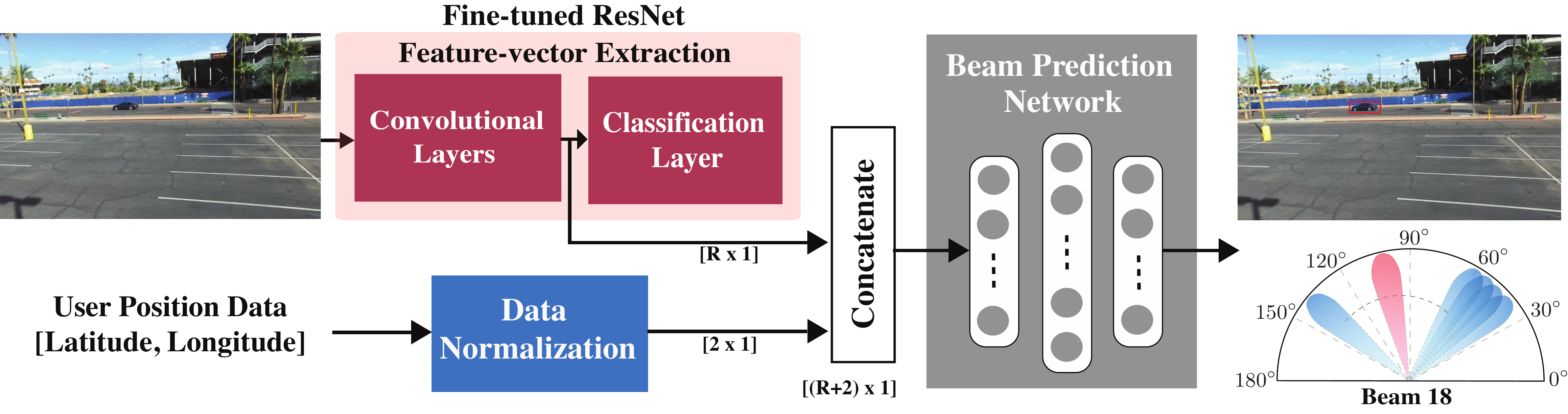}
	\caption{This figure illustrates the proposed multi-modal machine learning based beam prediction model that leverages both visual and (not necessarily accurate) position data for mmWave/THz beam prediction. }
	\label{main_fig2}
	\vspace{-4mm}
\end{figure*}

\section{Proposed Solution: Multi-Modal Vision-Position Based Beam Prediction} \label{sec:prop_sol}

In this paper, we propose a machine learning-based multi-modal beam prediction solution for the scenarios with a single candidate user in the visual scene. The multi-modality of data is associated with it own sets of challenges related to data fusion strategy, neural network design, etc. \cite{Lahat2015}. It is important to understand the relation between the different data modalities to be able to design an effective solution for the task of interest. In this section, we first present the reasoning behind choosing visual and position data for beam prediction and then present the details of the proposed machine learning solution.

\subsection{Key Idea}
The mmWave communication systems require large antenna arrays and use directive narrow beams to achieve high received signal power gain. This is primarily to overcome the severe path-loss associated with the high-frequency signals. The best narrow beamforming vector is selected from the pre-defined codebook $\boldsymbol{\mathcal F}$, which is typically associated with high training overhead. In general, the mechanism of directing the narrow beams can be viewed as focusing the signal into a particular direction in space. The beam vectors divide the scene (spatial dimensions) into multiple (possible overlapping) sectors, where each sector is associated with a particular beam value. Therefore, given a pre-defined codebook, the beam prediction task can be transformed into a classification task, where depending on where the user exists in the visual scene, a beam index from the codebook is assigned. 

With the recent advances in modern object detection models, it is possible to identify and extract the relative position of the user in a visual scene. Furthermore,  the advances in positioning systems such as GPS, the absolute position of any user in the scene can be obtained with some accuracy. Therefore, the location information of a user in the environment can be derived either from an image or from the position data. In this paper, instead of performing conventional beam training, we select the optimal beam index relying on the visual and/or position data. Here, it is important to evaluate the potential gains of leveraging both visual and position data compared to relying on only one of them. In the next subsection, we describe the proposed machine learning-based beam prediction model using the multi-modal visual and position data before delving into evaluating its performance in sections \ref{sec:datset} $-$ \ref{sec:perf_eval}.

\subsection{Machine Learning Model}\label{subsec:ml_model}
In this subsection, we propose three models to perform beam prediction: two different models for beam prediction using vision and position data separately and the third model, where both the modalities are merged and processed to predict the optimal beam index.  

\textbf{Vision-Based Beam Prediction:}\label{img_model}
As described in Section~\ref{sec:prob_form}, in this sub-task, the objective is to learn the prediction function $f_{\Theta}(\mathcal S)$ utilizing only visual data, i.e., RGB images. The proposed idea utilizes the state-of-the-art convolutional neural network (CNN) to perform this classification task, i.e., the model learns to map an image to a beam index. The primary choice of CNN for this particular task is an ImageNet2012 \cite{imagenet} pre-trained ResNet-50 \cite{resnet} model. The model is further modified by removing the last layer and replacing it with a fully-connected layer of size $M$ neurons. The fully-connected layer parameters are initialized randomly following a normal distribution with zero mean and unit variance. Different from conventional transfer learning \cite{NIPS2014_375c7134},
the ResNet-50 model is fine-tuned end-to-end in a supervised fashion, utilizing the labeled dataset described in \sref{sec:datset}.

\begin{figure*}[!t]
	\centering
	\subfigure[ ]{\includegraphics[width=0.315\linewidth]{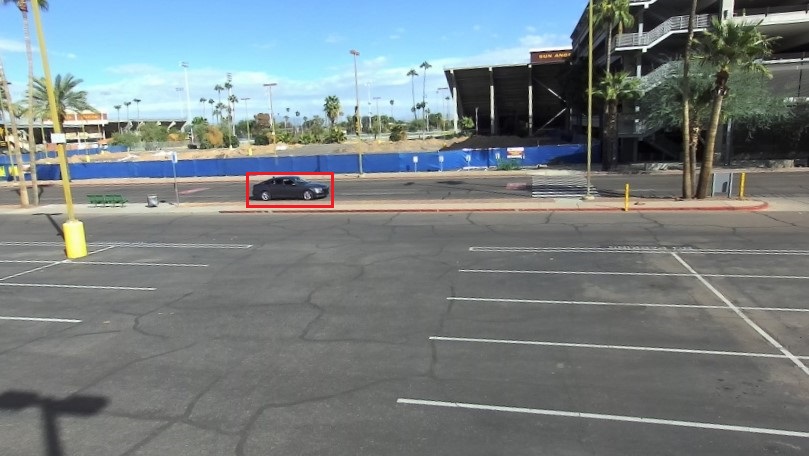}}
	\hspace{0.01\linewidth}
	\subfigure[ ]{\includegraphics[width=0.315\linewidth]{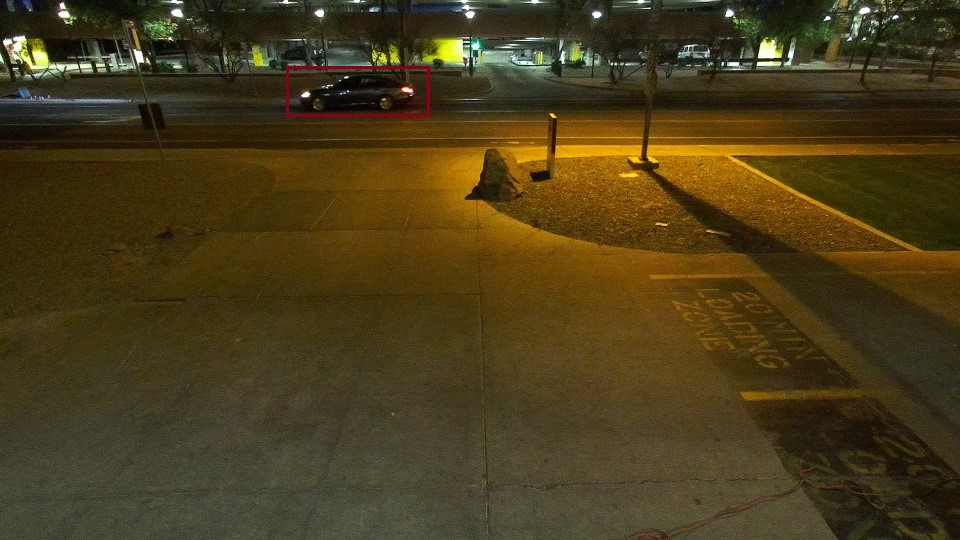}}
	\hspace{0.01\linewidth}
	\subfigure[ ]{\includegraphics[width=0.315\linewidth]{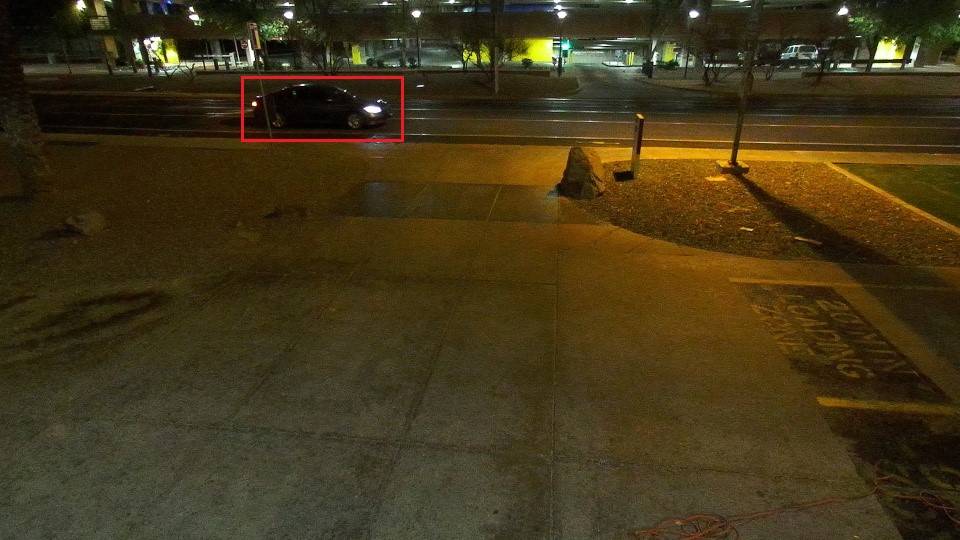}}
	\caption{The two wireless environments utilized in the data collection process is shown above. The transmitter, which is a vehicle in our dataset is highlighted in the above images. (a) shows the location and the data collected during day time, while (b) and (c) present the images captured during the night.  }
	\label{ref:data_samples}
	\vspace{-4mm}
\end{figure*}

\textbf{Position-based Beam Prediction:}\label{pos_model}
The goal of this sub-task is to predict the optimal beam index using the position information of the transmitter. The position information comprises of the Latitude and Longitude values as mentioned in Section~\ref{sec:prob_form}. The classification network for this sub-task is a Multi-Layer Perceptron (MLP) network with two hidden layers, each with $256$ hidden-units and an output layer with $M$ units. The inputs to the MLP network are the normalized values of Latitude and Longitude. ReLU activation function was applied to the output of the two hidden layers, whereas softmax function is applied to the final output layer.    

\textbf{Multi-modal Beam Prediction:}
In this sub-task, the optimal beam index is predicted based on both the visual and the position data. In order to efficiently merge the two modalities, we propose the neural network architecture in \figref{main_fig2}, which consists of two sequential stages: (i) The CNN-based feature extraction model, described in Section~\ref{img_model}, and (ii) a classifier MLP network. In the first stage, similar to the vision-based beam prediction, a ResNet 50 model is trained on the labeled images to predict the beam indices. The fine-tuned ResNet model is incorporated to the architecture after stripping off the final classifier layer. The objective is to extract the rich feature vector, $\textbf{V} \in \mathbb{R}^{T \times 1}$ from the convolutional layers. The position data (after normalization) is then concatenated to the extracted feature vector $\textbf{V}$, resulting in a combined vector $\widetilde{\textbf{V}} \in \mathbb{R}^{(T+2) \times 1}$. This combined vector $\widetilde{\textbf{V}}$ is provided as an input to the second stage of the architecture--- a two hidden-layered MLP network. The output of the MLP network produces a probability distribution $p \in \mathbb{R}^{M \times 1}$ over $M$ outputs, where $M$ is the total number of beams in the codebook.

\section{Testbed Description and Development Dataset}\label{sec:datset}
To accurately evaluate the performance of the proposed beam prediction approach in a realistic environment, we built a real-wold multi-modal dataset with vision, wireless, and position data. In this section, we first briefly describe the testbed used for the real wireless communication data collection and then present the analysis of the final development dataset used for the multi-modal beam-prediction task. 

\subsection{Data Collection Testbed}  \label{ssec:testbed}
We built a multi-modal mmWave communication testbed that consists of (i) a mobile transmitter (vehicle), (ii) a fixed receiver (acting as the base station), (iii) a position sensor at the vehicle (GPS antenna/receiver), and (iv) a visual data sensor (RGB camera) at the base station. The mobile transmitter has a 60GHz quasi-omni antenna and is always oriented towards the base station. The base station receiver has a 16-element 60GHz phased array adopting a beamforming codebook with $64$ beams. The RGB camera captures images of the wireless environment in the field-of-view of the base station. It is important to note here that the camera does not capture depth information (hence our approach relies on low-cost RGB cameras). The mmWave phased array and the camera are placed such that their fields of view are aligned. The base station is controlled by a laptop that operates the phased array and the camera. The transmitter is initialized by the base station laptop and works in a self-supervised fashion. The GPS receiver at the mobile transmitter collects position data that are then aligned with the RGB and wireless data. For further details, please refer to the data collection testbed description in the DeepSense 6G dataset \cite{ViWiReal}.

\begin{table}[!t]
	\caption{Design and Training Hyper-parameters}
	\centering
	\setlength{\tabcolsep}{5pt}
	\renewcommand{\arraystretch}{1.2}
	\begin{tabular}{@{}l|cc@{}}
		\toprule
		\toprule
		\textbf{Parameters}                     & \textbf{ResNet-50}  & \textbf{MLP}        \\ \midrule \midrule
		\textbf{Batch Size}                     & 32                  & 32                  \\
		\textbf{Learning Rate}                  & $1 \times 10 ^{-4}$ & $1 \times 10 ^{-2}$ \\
		\textbf{Learning Rate Decay}            & epochs 4 and 8      & epochs 20 and 40    \\
		\textbf{Learning Rate Reduction Factor} & 0.1                 & 0.1                 \\
		\textbf{Dropout}                        & 0.3                 & 0.3                 \\
		\textbf{Total Training Epochs}          & 15                  & 50                  \\ \bottomrule \bottomrule
	\end{tabular}
	\label{tab_nn_train_params}
	\vspace{-4mm}
\end{table}

\subsection{Development Dataset} \label{ssec:dataset}

In order to evaluate the performance of the proposed solution, two scenarios, namely, scenarios $5$ and $6$, were adopted from the DeepSense 6G dataset \cite{ViWiReal}. Both scenarios consist of diverse wireless, visual, and position data. Further, the two measurements of the two scenarios are collected at two different locations and at different times of the day (to have various lighting conditions), as presented in Fig.~\ref{ref:data_samples}. This ensures some diversity in the final development dataset. This diversity is further enhanced by adopting different base station heights and different distances between the base station and mobile user in the two scenarios (as described in \cite{ViWiReal}). Scenarios $5$ and $6$ consist of data collected during the night and day time, respectively.

The data collected from the two locations are utilized to create the development dataset for the multi-modal beam-prediction task. The initial data collected using the testbed is referred to as the raw data. The raw data further undergoes a post-processing pipeline. The first step involves downsampling the power vector, which comprises the beam powers of the $64$ beams, to $32$ elements by selecting every other element in the vector. Since the original $64$ beams are highly-overlapped, the down-sampling does not affect the total area covered by the beams. The index of the beam with the maximum received power is then selected as the optimal beam index. The resulting development dataset consists of $3462$ samples, with $2451$ and $1011$ samples in scenarios $5$ and $6$, respectively. The data from each location are further divided into training and validation sets with a split of $70 - 30\%$.

\section{Experimental Setup}\label{sec:exp_set}

The following two subsections discuss the neural network training parameters and the adopted evaluation metrics. 

\subsection{Network Training}
To highlight the impact of the different sensory data modalities on beam prediction task, three different machine learning models were proposed in Section~\ref{subsec:ml_model}, i.e., (i) vision-based beam prediction model, (ii) position-based beam prediction model, and (iii) multi-modal vision-position based beam prediction model. For the first two cases, the respective machine learning models, i.e., ResNet-50 and 2-layered MLP networks, are trained on the images and the position data of the training dataset. The multi-modal model is trained on the same training dataset using the feature vector extracted from the trained ResNet model and the ground-truth position values of the training dataset. The cross-entropy loss with Adam optimizer is used to train all three models. It is important to note here that for the two different scenarios mentioned in Section~\ref{sec:datset}, separate models were trained. The detailed hyperparameters used to fine-tune the models are presented in Table~\ref{tab_nn_train_params}. 

\subsection{Evaluation Metrics}
Now, we present the evaluation metric adopted to evaluate the performance of our proposed machine learning framework. Post-training, we evaluate each network on the validation set. The primary evaluation metric is top-1 accuracy. Top-2 and top-3 accuracies of the predicting model are also calculated to further examine the neural network's ability in beam prediction. Note that the top-k accuracy is defined as the percentage of the validation samples where the optimal ground-truth beam was within the top-$k$ predicted beams (i.e., the $k$ beam indices with the highest predicted probabilities).

\section{Performance Evaluation} \label{sec:perf_eval}

In order to develop an accurate evaluation of our proposed multi-modal beam prediction approach,  we first train and test the three developed models discussed in \sref{subsec:ml_model}, namely position-only, vision-only, and joint vision-position based beam prediction models. In Table~\ref{tab_acc}, the beam prediction accuracies (top-1, top-2, and top-3) of the three proposed models are presented for the two datasets/locations.
First, the results in this table show that the vision-based model achieves better beam prediction accuracies compared to the position-based solution for both the day and night datasets. This is mainly because the position data, collected by practical GPS sensors, are not accurate enough for narrow beam prediction. 
Table~\ref{tab_acc} also shows that the proposed multi-modal beam prediction model that leverages both the position and visual data can generally lead to higher beam prediction accuracies compared to the position-only and vision-only approaches. \textbf{It achieves more than 75\%  top-1 beam prediction accuracy and close to 100\% accuracy on top-3.} Note that, practically, 100\% top-3 accuracy means that performing over-the-air beam training for only the 3 most promising beams suggested by the proposed machine learning model leads to the optimal beam selection, i.e., yielding the same performance of the exhaustive search over the 32 beams. It is important to highlight here that the position-only beam prediction accuracy for the night dataset is lower than that of the day dataset. This can be attributed to the difference in the accuracy of the day and night position data in the DeepSense 6G dataset.

\begin{figure}[t]
	\centering
	\includegraphics[width=.9\columnwidth]{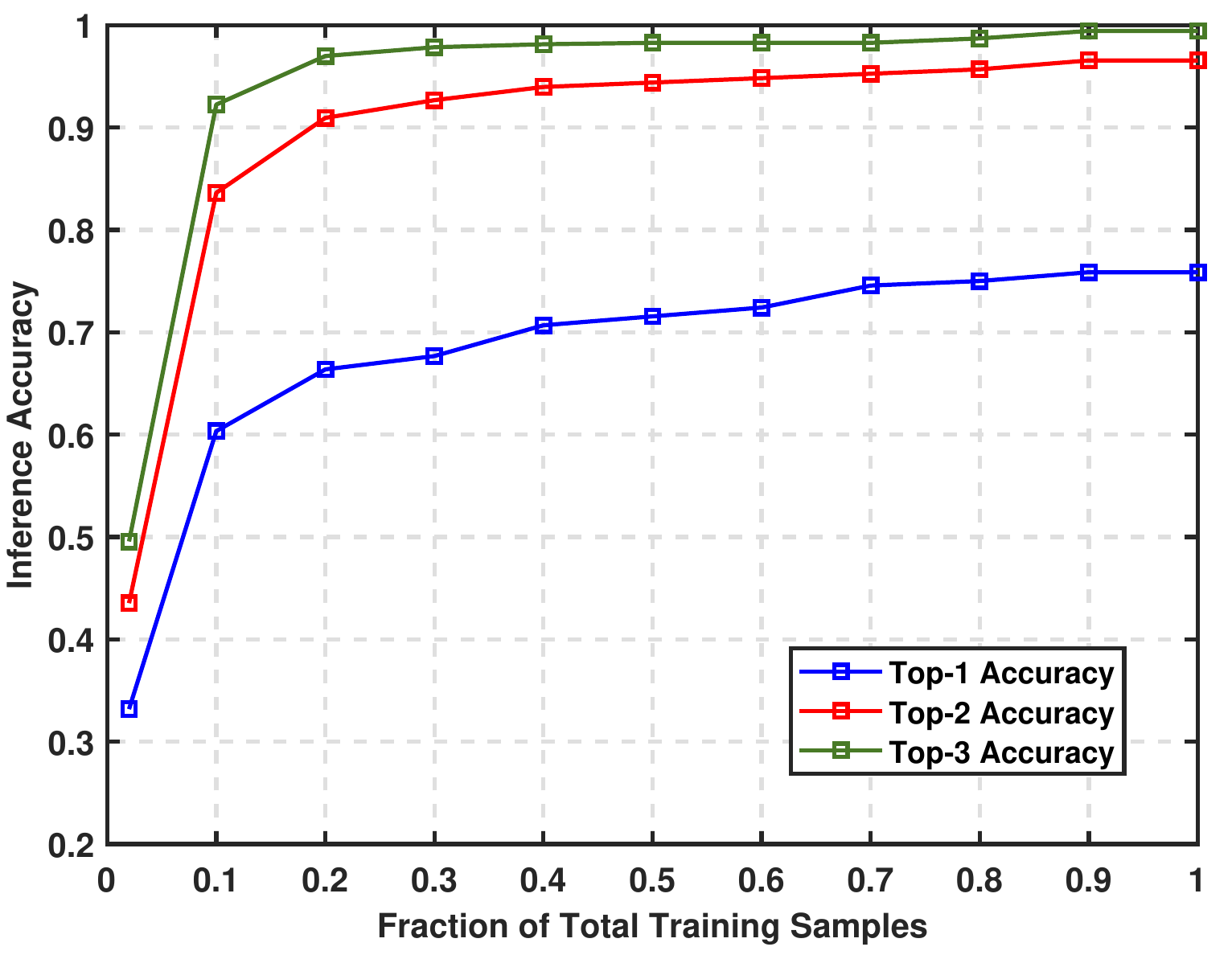}
	\caption{This figure shows the top-1, top-2, and top-3 beam prediction accuracy for the proposed vision-position model versus different dataset sizes.}
	\label{acc_vs_frac}
	\vspace{-4mm}
\end{figure}

Next, we try to draw some insights into the required dataset size. To do that, we present in \figref{acc_vs_frac} the top-1, top-2, and top-3 beam prediction accuracies versus the fraction of total training dataset size. Note that the total size of the real-world dataset described in \sref{ssec:dataset} is around 3000 data points divided into 70\% training and 30\% validation.  First, \figref{acc_vs_frac}  shows that the proposed multi-modal machine learning model is capable of learning the given beam prediction task with just $40\%$ of the total training samples, i.e., around 840 samples out of 2100 training samples.  Further, the figure illustrates that even smaller datasets may be required to achieve close to 100\% beam prediction accuracies when considering the top-3 predicted beams. These results highlight the efficiency and potential of the proposed multi-modal approach for the mmWave/THz beam prediction task.


\begin{table}[!t]
	\caption{Prediction Accuracies for the Proposed Models}
	\centering
	\setlength{\tabcolsep}{5pt}
	\renewcommand{\arraystretch}{1.2}
	\begin{tabular}{@{}l|ccc|ccc@{}}
		\toprule \toprule
		\multirow{2}{*}{\textbf{Model}} & \multicolumn{3}{c}{\textbf{Day Data Accuracy}}   & \multicolumn{3}{c}{\textbf{Night Data Accuracy}} \\ \cmidrule(l){2-7} 
		& \textbf{Top-1} & \textbf{Top-2} & \textbf{Top-3} & \textbf{Top-1} & \textbf{Top-2} & \textbf{Top-3} \\ \midrule \midrule
		\textbf{Vision-only}            & 0.7127         & 0.9482         & 0.9913         & 0.7225         & 0.9531         & 0.9903         \\ 
		\textbf{Position-only}          & 0.6163         & 0.8965         & 0.9741         & 0.5331         & 0.7876         & 0.8933         \\ 
		\textbf{Vision-Position}        & \textbf{0.7586} & \textbf{0.9655} & \textbf{0.9955} & \textbf{0.7371} & \textbf{0.9634} & \textbf{0.9919} \\ \bottomrule \bottomrule
	\end{tabular}
	\label{tab_acc}
\end{table}


\section{Conclusion}\label{sec:conc}
This paper develops a novel approach that leverages multi-modal machine learning and visual/positional sensory data for fast mmWave/THz beam prediction. To evaluate the proposed approach, we constructed a real-world multi-modal vehicular dataset that comprises position, vision (camera), and mmWave beam training data. Using this realistic dataset, our developed model achieved more than 75\% success probability in predicting the optimal beam without any beam training overhead. Further, it achieved close to 100\% top-3 beam prediction accuracy. This highlights the promising gains of leveraging multi-modal data for significantly reducing the beam training overhead in mmWave/THz communication systems.

\balance

\end{document}